\documentclass[prl,aps,showpacs,twocolumn,superscriptaddress]{revtex4}
\usepackage{graphics,bm}
\usepackage{amssymb,ulem}
\usepackage{epsfig}
\usepackage{epsf}
\usepackage[usenames]{color}

\begin{document}
\title{Absence of Wigner molecules in one-dimensional few-fermion systems with short-range interactions}
\author{Jing-Jing Wang}
\affiliation{Department of Physics, Zhejiang Normal University, Jinhua, Zhejiang Province, 321004, China}
\author{Wei Li}
\affiliation{Department of Physics, Fudan University, Shanghai, 200433, China}
\author{Shu Chen}
\affiliation{Beijing National Laboratory for Condensed Matter Physics, Institute of Physics, Chinese Academy of Sciences, Beijing 100190, China}
\author{Gao Xianlong}
\email{gaoxl@zjnu.edu.cn}
\affiliation{Department of Physics, Zhejiang Normal University, Jinhua, Zhejiang Province, 321004, China}
\author{Massimo Rontani}
\email{massimo.rontani@nano.cnr.it}
\affiliation{S3, Istituto Nanoscienze-CNR, Via Campi 213/A, 41125 Modena, Italy}
\author{Marco Polini}
\affiliation{NEST, Istituto Nanoscienze-CNR and Scuola Normale Superiore, 56126 Pisa, Italy}

\begin{abstract}
We study by means of exact-diagonalization techniques the ground state of a few-fermion system with strong short-range repulsive interactions trapped by a harmonic potential in one spatial dimension. Even when the ground-state density profile displays at strong coupling very well pronounced Friedel oscillations with a `$4k_{\rm F}$ periodicity', the pair correlation function does not show any signature of Wigner-molecule-type correlations. For the sake of comparison, we present also numerical results for few-electron systems with Coulomb interactions, demonstrating that their ground state at strong coupling is, on the contrary, a Wigner molecule.
\end{abstract}

\pacs{03.75.Ss, 67.85.Lm, 71.10.Pm, 73.20.Qt}

\maketitle

{\it Introduction.} --- Understanding the nature of the ground state of strongly correlated 
few-fermion systems has attracted a great deal of interest. 
A large body of literature is available in the context of 
few-electron quantum dots, or `artificial atoms'~\cite{generalqds}. 
In this case the relevant interaction potential is the long-range Coulomb force 
between electrons. It is by now well established that when Coulomb 
interactions 
are sufficiently strong the ground-state of a 
quantum dot --- either one-~\cite{wignermolecules1Dquantumdots} or two-dimensional~\cite{wignermolecules2Dquantumdots} --- 
is a `Wigner molecule' (WM). This jargon stems from the Wigner crystalline order that is 
displayed by an electron gas at ultra-low 
densities~\cite{Giuliani_and_Vignale}. 
The WM state is not characterized by a spontaneous breaking of 
translational symmetry (which is absent to begin with due to the confining potential of the dot) 
but rather by strong short-range order that is evident in the density-density 
correlation function (the so-called pair correlation function). 
The existence of WMs has been 
recently confirmed by inelastic light 
scattering studies of few-electron circular 
quantum dots~\cite{kalliakos_naturephys_2008} as well as by tunneling
spectroscopies of quantum wires~\cite{WMwires} and carbon nanotubes~\cite{WMnanotubes}.

The ground state of one-dimensional (1D) systems of fermions 
trapped in a parabolic potential and interacting through a 
{\it short-range} potential 
has also been the subject of numerous analytical and numerical studies~\cite{1DFermigasestheory,BALDA}. The interest in these systems is not merely academic since gases of fermionic atoms trapped in `atomic quantum wires' can be realized experimentally~\cite{moritz_prl_2005} and cooled to temperatures $T \sim 0.1~T_{\rm F}$, where $T_{\rm F}$ is the Fermi temperature~\cite{liao_naturephys_2010}. The relevant interaction potential in this case is a zero-range Fermi pseudopotential acting only between antiparallel-spin fermions~\cite{footnote-spin}, $V(x) = g_{\rm 1D}\delta(x)$, with an amplitude $g_{\rm 1D}$ which is controlled~\cite{olshanii_prl_1998} by the three-dimensional (3D) scattering length. 
Recent advances in atomic physics have made possible also to confine only a few fermionic atoms with tunable short-range interactions in optical dipole traps~\cite{serwane_science_2011}. 

In the non-interacting $g_{\rm 1D} \to 0$ limit the 
ground-state density profile of $N$ fermions trapped in 
a parabolic potential displays, despite the smooth 
boundaries, `Friedel oscillations' with $N/2$ peaks~\cite{noninteracting_1D}, which are ultimately due to the Pauli principle. In the non-interacting limit, indeed, the ground-state density $n_0(x)$ is obtained by occupying exactly $N/2$ harmonic-oscillator levels, each one with two fermions with antiparallel spin: $n_0(x) = 2 \sum_{n=0}^{N/2-1}|\phi_n(x)|^2$, where $\phi_n(x) = (2^n n! \pi^{1/2} \ell_{\rm ho})^{-1/2}\exp{(-\xi^2/2)}~H_n(\xi)$ with $\xi = x/\ell_{\rm ho}$ and $\ell_{\rm ho} = (\hbar/m\omega)^{1/2}$ 
are the eigenfunctions of a 1D harmonic oscillator with frequency $\omega$. Here $H_n(\xi)$ is a Hermite polynomial of degree $n$. Straightforward mathematical manipulations yield the following approximate expression for $n_0(x)$ away from the trap edges~\cite{soffing_pra_2011}:
\begin{equation}
n_0(x) \approx n_{\rm TF}(x) - \frac{(-1)^{N/2}}{\pi L_{\rm TF}}\frac{\cos[2k_{\rm F}(x)x]}{1- x^2/L^2_{\rm TF}}~,
\end{equation}
where 
$n_{\rm TF}(x) = (2 L_{\rm TF}/\ell^2_{\rm ho})[1-(x/L_{\rm TF})^2]^{1/2}$ 
is the Thomas-Fermi density profile, $L_{\rm TF} = \ell_{\rm ho}N^{1/2}$, and $k_{\rm F}(0) =\pi n_{\rm TF}(0)/2 = N^{1/2}/\ell_{\rm ho}$ (the expression for $k_{\rm F}(x)$ away from the trap center can be found 
in Ref.~\onlinecite{soffing_pra_2011}).

What happens to this simple single-particle physics when the strength of inter-particle interactions is increased? In the limit $g_{\rm 1D} \to \infty$ the delta-function interaction imposes an effective Pauli principle between antiparallel-spin fermions. In this limit we thus expect oscillations in the ground-state density $n_\infty(x)$ with a bulk periodicity controlled by $4 k_{\rm F}(0)$. These have been dubbed in the literature `Wigner oscillations'. The transition from the $2k_{\rm F}(0)$ Friedel oscillations to the $4k_{\rm F}(0)$ Wigner oscillations in a parabolic trap is a smooth crossover: see, for example, the extensive density-matrix renormalization group study by S\"{o}ffing {\it et al.}~\cite{soffing_pra_2011}. 

Looking, however, at a one-body observable such as the ground-state density profile does not shed light on the nature of the ground state at strong coupling. Especially important in low-dimensional systems are indeed correlation functions. In the absence of the parabolic trapping, we know from bosonization~\cite{giamarchibook} that at strong coupling, {\it i.e.} when $K_\rho < 1/3$, $K_\rho$ being the Luttinger liquid parameter in the charge sector, 1D interacting fermions are dominated by $4k_{\rm F}$ charge-density-wave (CDW) correlations. Since in a Galilean invariant system $K_\rho = v_{\rm F}/ v_{\rho}$, where $v_{\rm F}$ is the bare Fermi velocity and $v_\rho$ is the velocity of density excitations, and $v_\rho \to 2 v_{\rm F}$ in the limit $g_{\rm 1D} \to \infty$~\cite{1DFermigasestheory}, we expect that the ground state at strong coupling is completely dominated by $2k_{\rm F}$ spin-density-wave correlations with no space for incipient CDWs. The situation changes dramatically in a system with long-range interactions, which stabilize $4k_{\rm F}$ CDW correlations~\cite{glazman_prb_1992,schulz_prl_1993,giamarchibook}. 

In this Letter we analyze the nature of the ground state of a 1D few-fermion system at strong coupling in the presence of a parabolic potential, which breaks Galilean invariance. More precisely, inspired by the body of literature discussed above, we study both ground-state density profiles {\it and} pair correlation functions. We focus on the strong coupling regime, which is difficult to access with analytical techniques, and rely on accurate numerical calculations based on the exact-diagonalization method (aka `full configuration 
interaction')~\cite{rontani_jcp_2006}. We find that even when the ground-state density profile displays well pronounced Wigner oscillations with a $4k_{\rm F}(0)$ periodicity, the pair correlation function does not show any signature of WM-type correlations. We then highlight the role of the interaction range by comparing these findings with those for a system of electrons interacting through the Coulomb potential. 
We demonstrate indeed that the ground state of this system at strong coupling is a WM. 

{\it The model and the observables of interest.} ---  We consider a two-component Fermi gas with $N$ atoms confined inside a strongly elongated harmonic trap.
The two species of fermionic atoms are assumed to have the same mass $m$ and different pseudospin $\sigma=\uparrow$ or $\downarrow$ (hyperfine-state label). The trapping potential is axially symmetric and characterized by angular frequencies $\omega_\perp$ and $\omega$ in the radial and longitudinal directions, respectively, with $\omega \ll \omega_\perp$. Correspondingly, we introduce the harmonic-oscillator lengths $a_\perp=(\hbar/m\omega_\perp)^{1/2}$ and $\ell_{\rm ho}=(\hbar/m\omega)^{1/2}$.

The gas is dynamically 1D if the anisotropy parameter of the trap is much smaller than the inverse atom number, $\omega/\omega_\perp\ll N^{-1}$. It can thus be described by an inhomogeneous Gaudin-Yang Hamiltonian,
\begin{eqnarray}\label{eq:igy}
{\hat {\cal H}} &=& {\hat {\cal T}} + {\hat {\cal V}} + {\hat {\cal W}} = -\frac{\hbar^2}{2m}\sum_\sigma\int_{-\infty}^{+\infty}dx~{\hat \Psi}^\dagger_\sigma(x) \partial^2_x {\hat \Psi}_\sigma(x) \nonumber\\
&+& g_{\rm 1D}\int_{-\infty}^{+\infty}dx~{\hat \Psi}^\dagger_{\uparrow}(x){\hat \Psi}^\dagger_{\downarrow}(x){\hat \Psi}_{\downarrow}(x){\hat \Psi}_{\uparrow}(x)  \nonumber\\
&+& \frac{1}{2} m \omega^2\sum_{\sigma} \int_{-\infty}^{+\infty}dx~{\hat \Psi}^\dagger_\sigma(x) x^2 {\hat \Psi}_\sigma(x)~,
\end{eqnarray}
where ${\hat \Psi}^\dagger_\sigma(x)$ [${\hat \Psi}_\sigma(x)$] is a field operator that creates (destroys) a fermion with spin $\sigma$ at position $x$ and $g_{\rm 1D} \simeq 4\hbar^2 a_{\rm sc}/(m a^2_\perp)$ (in the limit $a_{\rm sc} \ll a_\perp$) is a parameter that determines the strength of inter-particle repulsions~\cite{olshanii_prl_1998}. The 3D scattering length $a_{\rm sc}$ can be tuned by means of a magnetic field~\cite{moritz_prl_2005}. The first term in Eq.~(\ref{eq:igy}) (${\hat {\cal T}}$) is the kinetic energy whereas the second term (${\hat {\cal V}}$) describes two-body short-range interactions. Finally, the third term (${\hat {\cal W}}$) is the parabolic trapping potential.

Choosing $\ell_{\rm ho}$ as unit of length and $\hbar\omega$ as unit of energy, 
the Hamiltonian ({\ref{eq:igy}) is governed by 
the dimensionless coupling parameter
\begin{equation}\label{eq_lambda}
\lambda=\frac{g_{\rm 1D}}{\ell_{\rm ho}\hbar\omega}~.
\end{equation} 
In this Letter we focus our attention on the dependence of two key quantities on $\lambda$ for strong inter-particle repulsions ($\lambda \gg 1$): i) the local spin-resolved density, $n_{\sigma}(x) = \langle 
{\hat \Psi}^\dagger_\sigma(x){\hat \Psi}_\sigma(x)\rangle/N_{\sigma}$,
and ii) the  pair correlation function (PCF)
\begin{equation}\label{eq:PCF}
{\bar g}_{\sigma\sigma'}(x) = 
\int_{-\infty}^{+\infty}dx'~g_{\sigma\sigma'}(x'+\frac{x}{2},x'-\frac{x}{2})~,
\end{equation}
which is defined in terms of the two-body correlator
\begin{equation}\label{eq:PCF}
g_{\sigma\sigma'}(x,x') = \frac{
\langle {\hat \Psi}^\dagger_{\sigma}(x){\hat \Psi}^\dagger_{\sigma'}(x'){\hat \Psi}_{\sigma'}(x'){\hat \Psi}_{\sigma}(x)\rangle
}{N_{\sigma} (N_{\sigma'} - \delta_{\sigma\sigma'}) }~,
\end{equation}
$N_{\sigma}$ being the number of atoms with spin $\sigma$.
We also introduce the total density, $n(x) = \sum_\sigma N_{\sigma} n_\sigma(x) / N$, 
and the total PCF, ${\bar g}(x) = 
\sum_{\sigma, \sigma'} 
N_{\sigma} (N_{\sigma'} - \delta_{\sigma\sigma'})
{\bar g}_{\sigma\sigma'}(x) / N(N-1)$. 
In all the definitions above $\langle \dots \rangle$ denotes the expectation value over the ground state of the Hamiltonian (\ref{eq:igy}). 
We remind that $g_{\sigma\sigma'}(x,x')$ measures the 
conditional probability of finding a fermion with spin $\sigma'$ 
at position $x'$ when another fermion of spin $\sigma$ is known to 
be at position $x$, whereas  ${\bar g}_{\sigma\sigma'}(x)$
provides the probability of finding two fermions at a relative distance $x$.

For the sake of comparison, we present also numerical results for a few-electron system in the same trapping potential. In this case the second quantized Hamiltonian is identical to the one in Eq.~(\ref{eq:igy}) except for the term in the second line, which now reads:
\begin{eqnarray}\label{eq:longrangeinteractions}
{\hat {\cal V}}_{\rm C} &=& \frac{1}{2}\sum_{\sigma,\sigma'}\int_{-\infty}^{+\infty}dx\int_{-\infty}^{+\infty}dx'~{\hat \Psi}^\dagger_\sigma(x){\hat \Psi}^\dagger_{\sigma'}(x')\nonumber\\
&\times&V_{\rm C}(|x-x'|){\hat \Psi}_{\sigma'}(x'){\hat \Psi}_\sigma(x)~.
\end{eqnarray}
As a model long-range potential we use the following regularized Coulomb interaction
\begin{equation}\label{eq:Coulomb}
V_{\rm C}(|x-x'|) = \frac{e^2}{\kappa\sqrt{a^2 + (x-x')^2}}~,
\end{equation}
where $a$ is a real positive quantity with physical dimensions of length
and $\kappa$ is the relative dielectric constant of the host semiconductor.  
The Hamiltonian ${\hat {\cal H}}_{\rm C} = {\hat {\cal T}} + {\hat {\cal V}}_{\rm C} + {\hat {\cal W}}$ describes few-electron quantum dots embedded 
e.g.~in a 1D quantum wire~\cite{wignermolecules1Dquantumdots}.
Using the same dimensionless variables which have been used above for the inhomogeneous Gaudin-Yang Hamiltonian, we find that the dimensionless coupling constant that determines the strength of electron-electron interactions is given by
\begin{equation}
\lambda_{\rm C}  = \frac{e^2}{\kappa\ell_{\rm ho} \hbar \omega}~.
\end{equation}
Due to the presence of the parameter $a$ in Eq.~(\ref{eq:Coulomb}), the few-electron problem does not depend only on $\lambda_{\rm C}$ but also on the ratio $a/\ell_{\rm ho}$, whose value has been fixed for all our numerical calculations to $a/\ell_{\rm ho} = 10^{-1}$ \cite{EDdetails}. 

{\it Numerical results.} --- In Fig.~\ref{fig:one} we report numerical results for the spin-resolved density profiles $n_{\sigma}(x)$ of $N=4$ and $N=5$ fermionic atoms.  
All the data presented in this figure and in all the figures below have been calculated for $\lambda =15$ and are qualitatively similar to those obtained
for other values of $N$, whereas $N>6$ is beyond our 
reach. In this ultra-strong coupling regime mean-field-like methods yield completely incorrect results (while we have checked that our numerical results at weak coupling can be explained very well by Bethe-Ansatz density-functional theory~\cite{BALDA}). In both panels of Fig.~\ref{fig:one} we clearly see $N$ distinct peaks in the total density profile $n(x)$, i.e., Wigner oscillations with a $4 k_{\rm F}(0)$ periodicity. Note also the `antiferromagnetic' spatial pattern of the spin-resolved densities $n_\sigma(x)$ in which $n_\downarrow(x)$ has a maximum in correspondence of every minimum of $n_\uparrow(x)$.
\begin{figure}[h!]
\begin{center}
\includegraphics[width=1.0\linewidth]{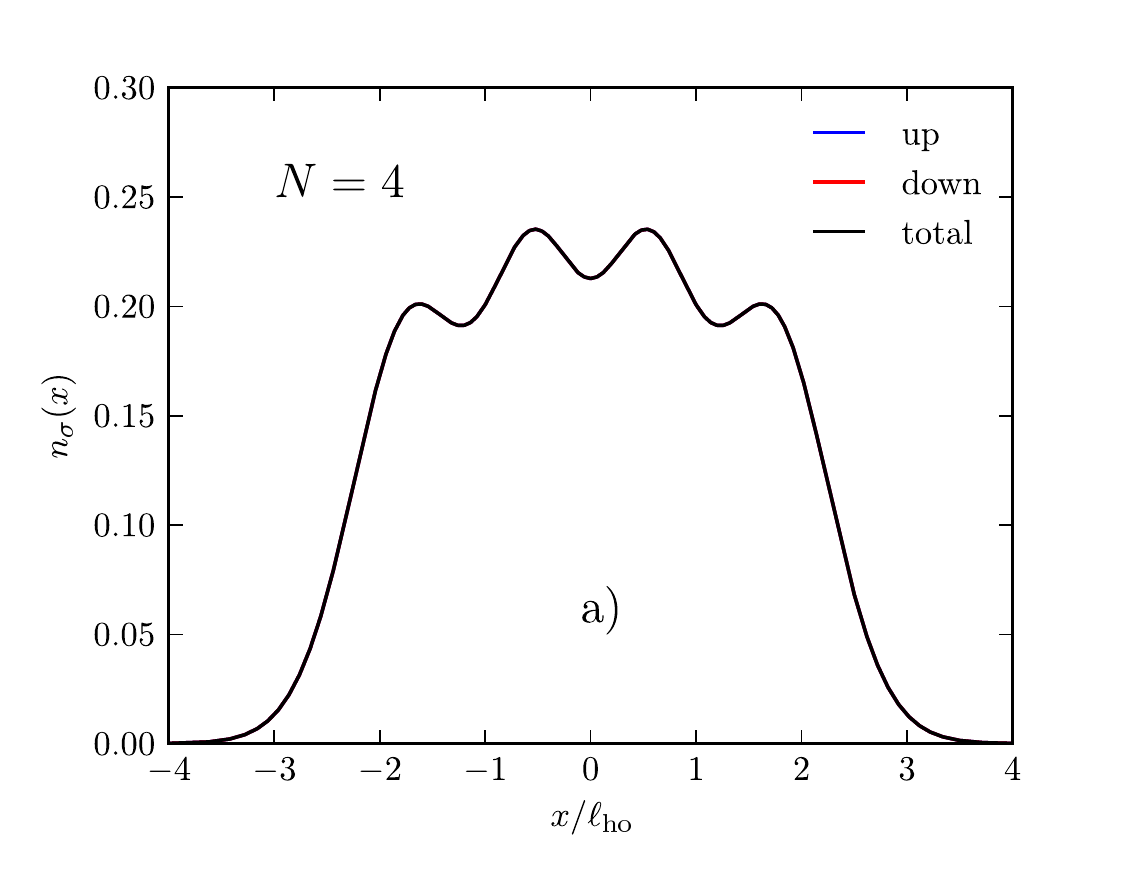}\\
\includegraphics[width=1.0\linewidth]{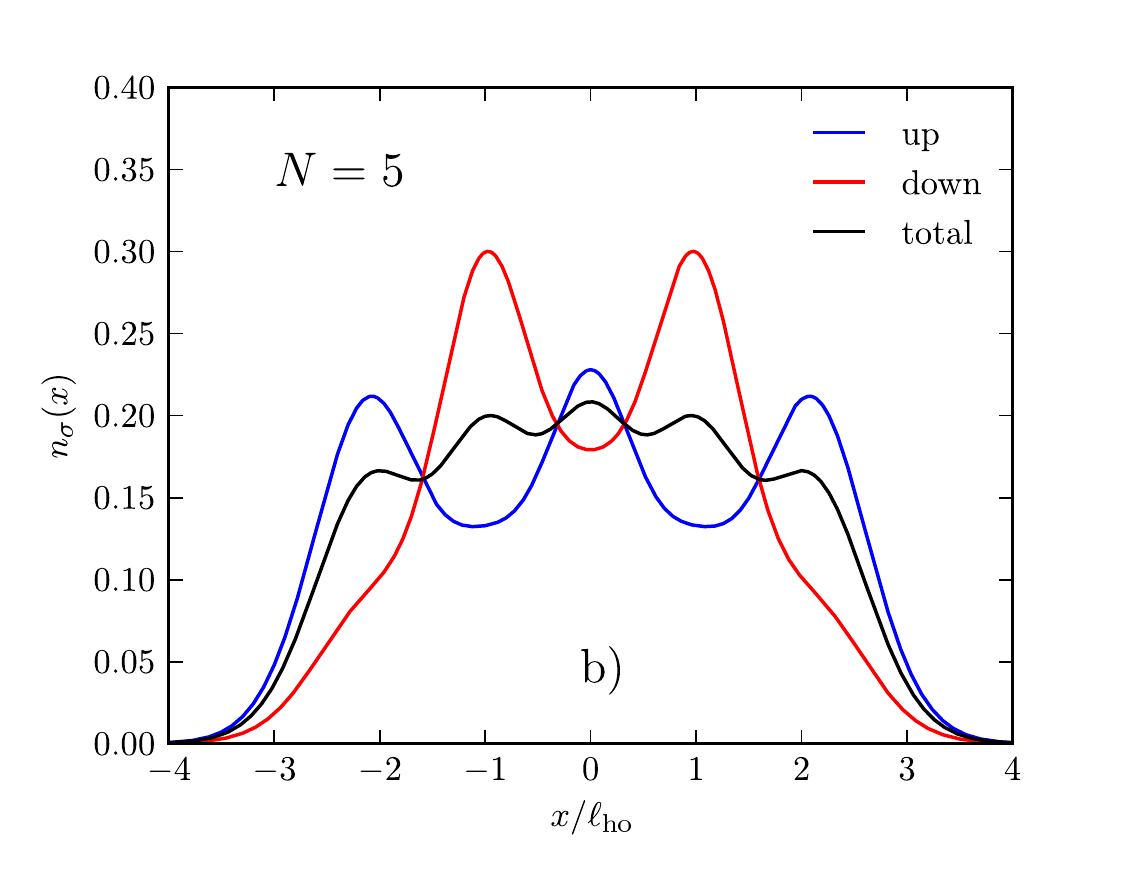}
\end{center}
\caption{(Color online)  Spin-resolved and total ground-state density profiles (in units of $\ell^{-1}_{\rm ho}$) as functions of 
$x/\ell_{\rm ho}$ for $\lambda =15$. Panel a) Results for $N = 4$ particles
(all curves overlap). Panel b) Results for $N = 5$. 
\label{fig:one}}
\end{figure}
Quite surprisingly, the PCFs corresponding to the density profiles in Fig.~\ref{fig:one}, which are plotted in Fig.~\ref{fig:two}, 
do {\it not} display any sign of WM-type short-range order. Despite the well-defined Wigner oscillations in the density profile, the ground state seems to display liquid-type correlations, 
without any sign of interaction-induced localization.
\begin{figure}[h!]
\begin{center}
\includegraphics[width=1.0\linewidth]{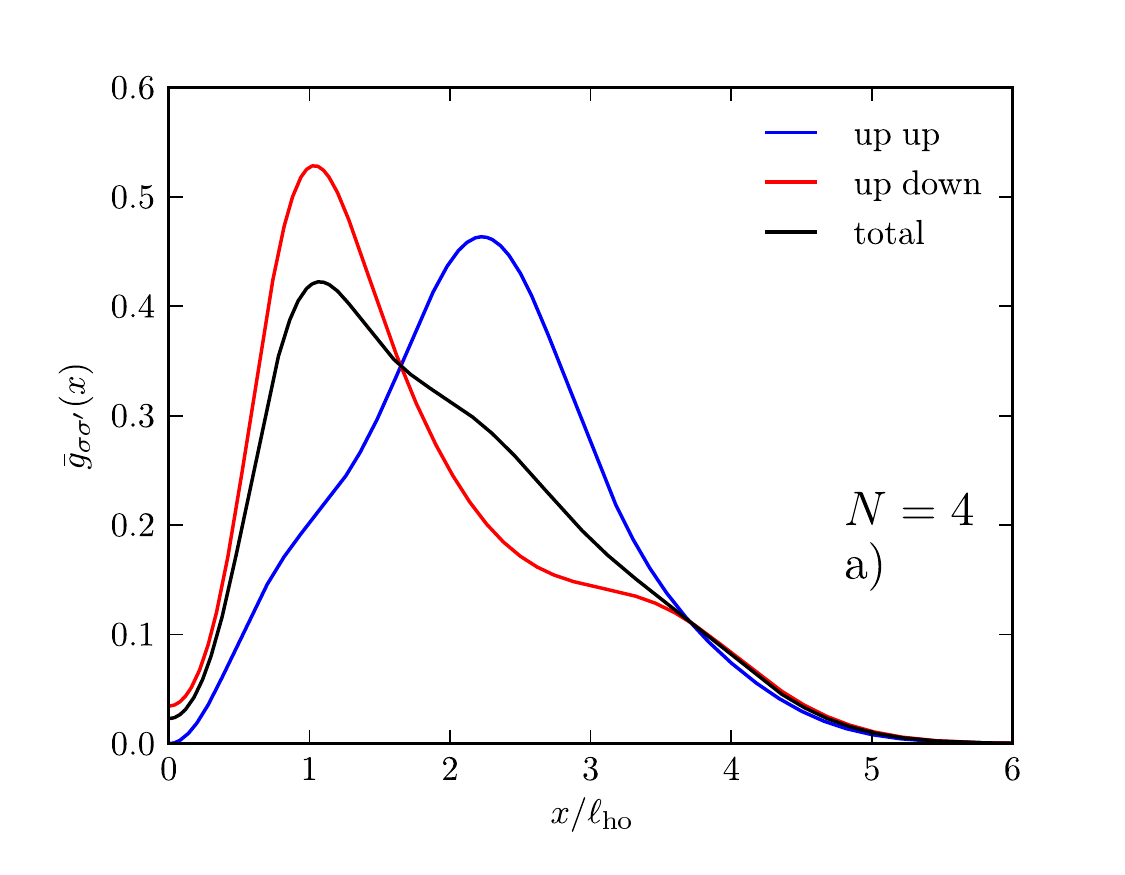}\\
\includegraphics[width=1.0\linewidth]{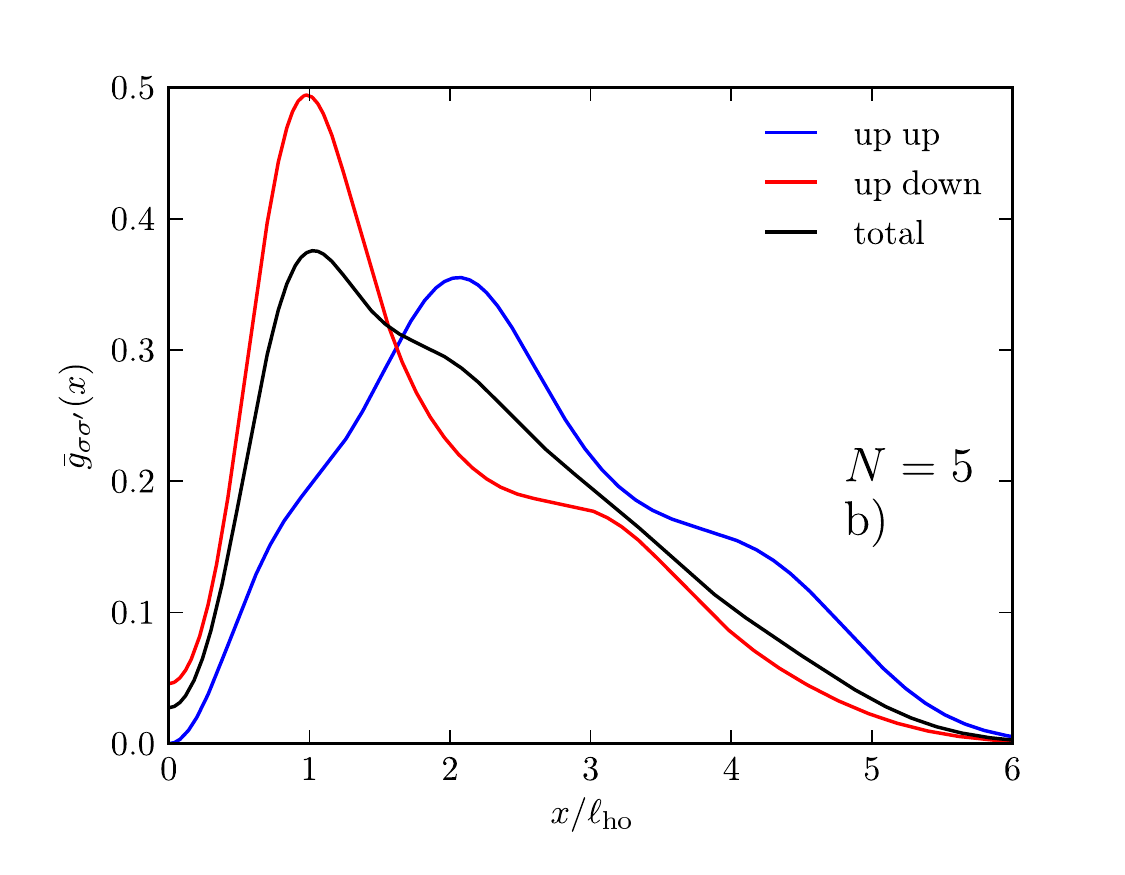}
\end{center}
\caption{(Color online)  Pair correlation function 
${\bar g}_{\sigma\sigma'}(x)$ (in units of $\ell^{-1}_{\rm ho}$)
vs $x/\ell_{\rm ho}$ for the same system parameters as in Fig.~\ref{fig:one}. 
Note that ${\bar g}_{\uparrow\uparrow}(x=0) = 0$, a manifestation of Pauli's exclusion principle. 
\label{fig:two}}
\end{figure}

In Fig.~\ref{fig:three} we compare the results reported in Fig.~\ref{fig:one} with those for few electrons interacting through the Coulomb potential. For the sake of comparison, in producing the data for Fig.~\ref{fig:three} we have chosen $\lambda_{\rm C} = 15$, which is the same value used for the short-range coupling $\lambda$ in Fig.~\ref{fig:one}. Also the total density profile in Fig.~\ref{fig:three} is characterized by Wigner oscillations with a contrast though which is much higher than in the case of fermions interacting through short-range forces. 
\begin{figure}
\begin{center}
\includegraphics[width=1.0\linewidth]{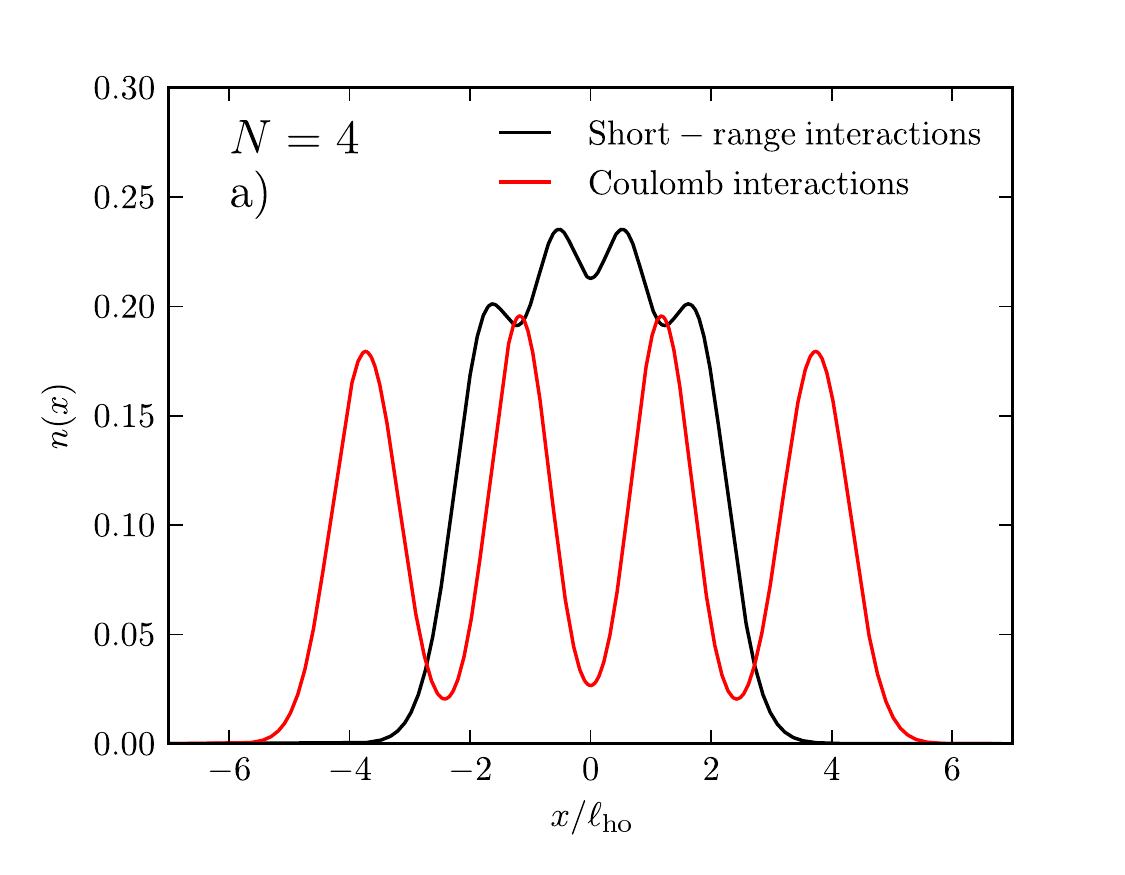}\\
\includegraphics[width=1.0\linewidth]{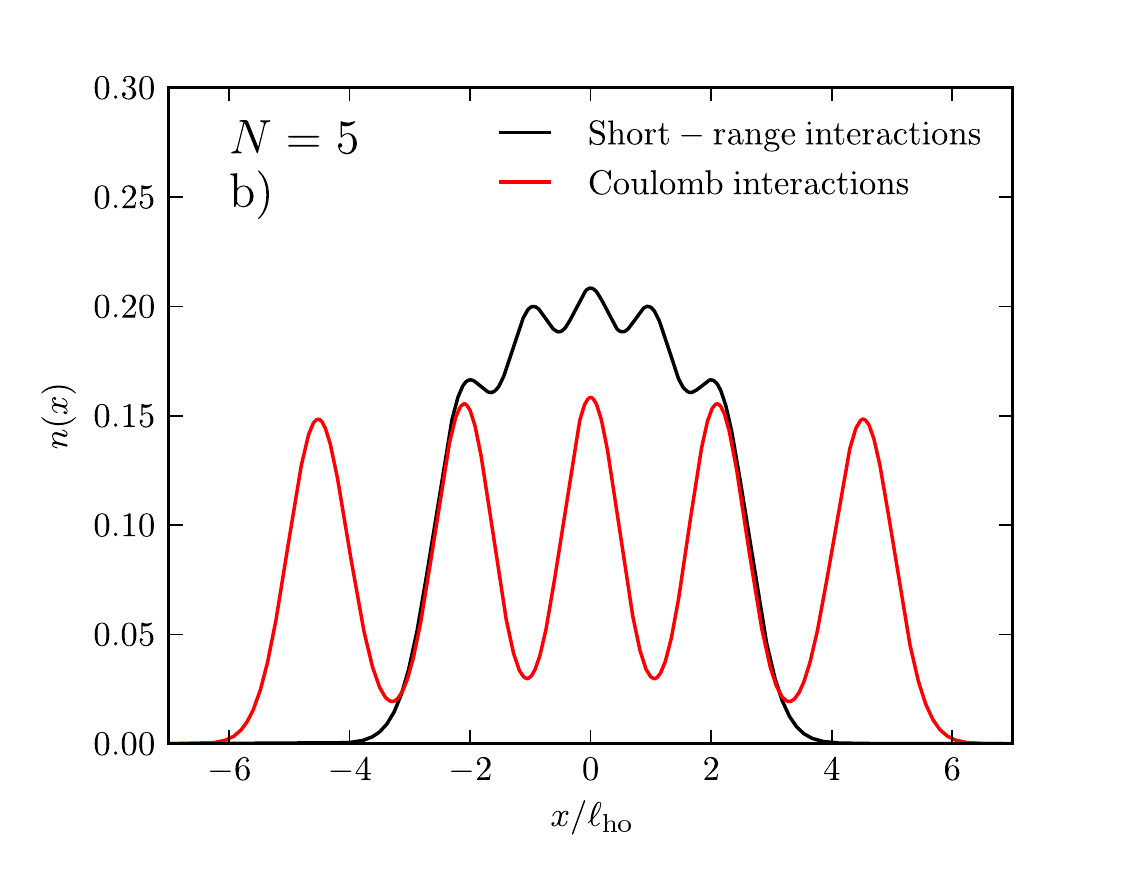}
\end{center}
\caption{(Color online)  The total density profile $n(x)$ of a system of fermions interacting with short-range interactions is compared with that of electrons interacting through Coulomb interactions. The data for short-range interactions have been produced by setting $\lambda = 15$ (same data shown in Fig.~\ref{fig:one}). The data for Coulomb interactions have been produced by setting $\lambda_{\rm C} =15$ and $a/\ell_{\rm ho} =10^{-1}$.\label{fig:three}}
\end{figure}
Moreover, at this large value of $\lambda_{\rm C}$, the ground state of the few-electron system is definitely a WM. This is highlighted in Fig.~\ref{fig:four} where we have presented the total PCFs corresponding to the density profiles in Fig.~\ref{fig:three}. In the Coulomb-coupled system we note an `excluded volume' region, i.e., a finite range of values of $x$ close to the origin in which ${\bar g}(x) \sim 0$, which is typical of electronic systems at ultrastrong coupling~\cite{gzeroCoulomb}. Note also that ${\bar g}(x)$ displays exactly $N-1$ high-contrast peaks, a clearcut signature of the WM nature of the ground state. 
\begin{figure}
\begin{center}
\includegraphics[width=1.0\linewidth]{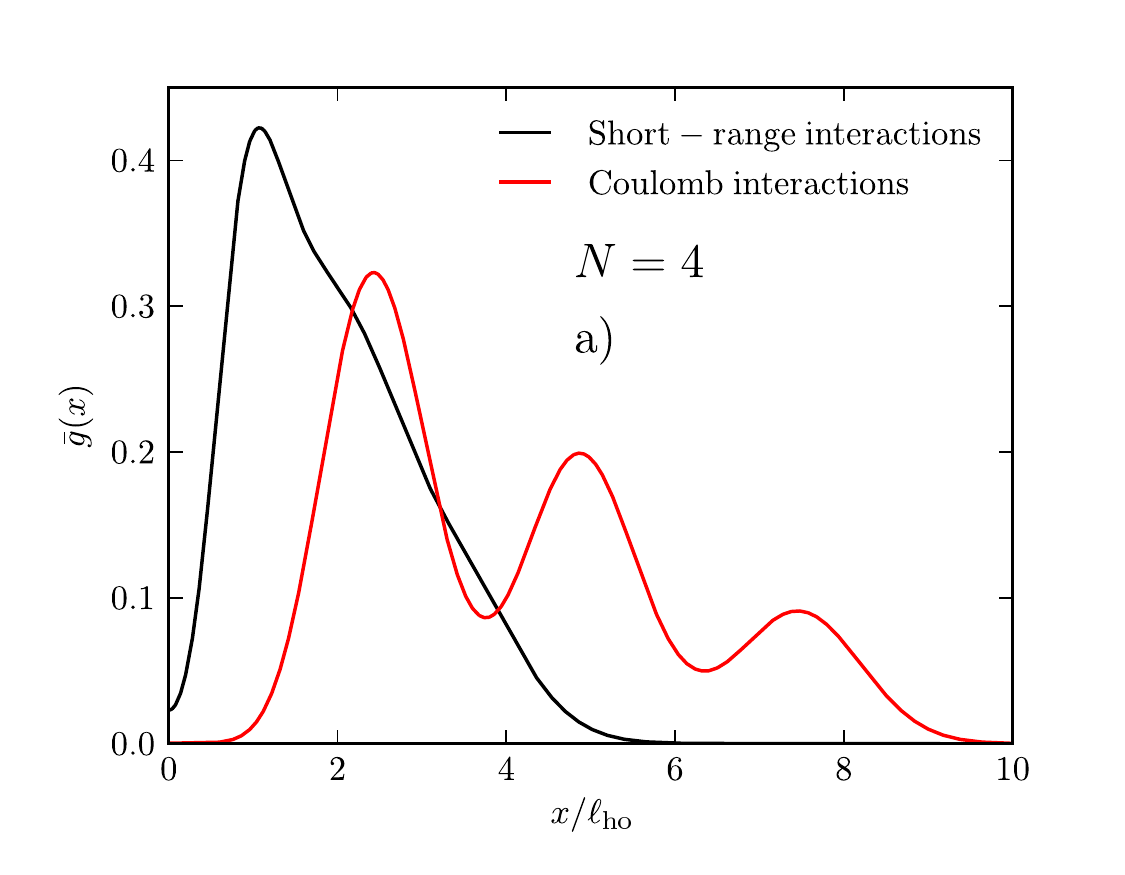}\\
\includegraphics[width=1.0\linewidth]{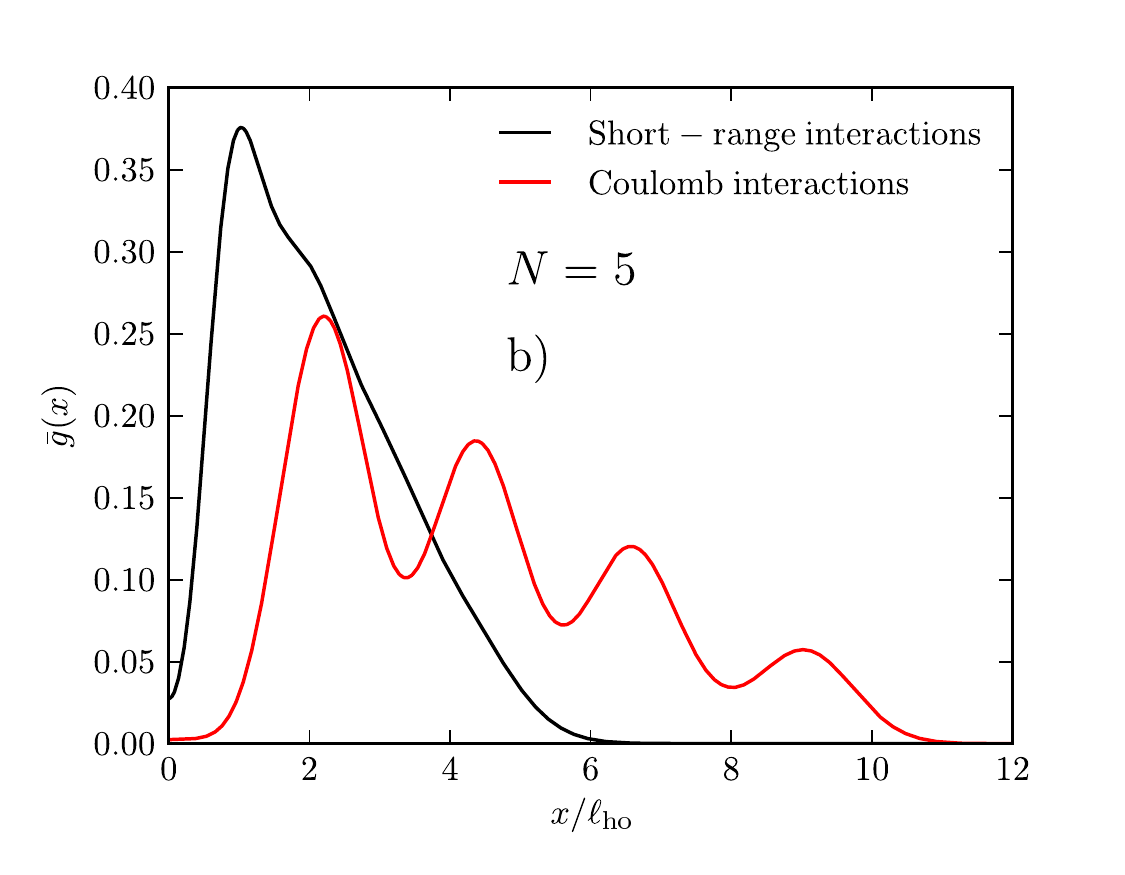}
\end{center}
\caption{(Color online) The total pair correlation function ${\bar g}(x)$ of a system of fermions interacting with short-range interactions is compared with that of electrons interacting through Coulomb interactions. The parameters used to produced the data in this figure are identical to those in Fig.~\ref{fig:three}.\label{fig:four}}
\end{figure}

In conclusion, we have studied density profiles and pair correlation functions of few-fermion systems with short-range interactions at strong coupling. We have discovered that the nature of the ground state of these systems, at least in the regime of coupling constants we have investigated, is not a Wigner molecule, despite the well defined `Wigner oscillations' displayed by the density profile. We believe that the ground state of this system is not `localized' into a Wigner molecule because of the short-range nature of the inter-atom interactions. Systems with identical statistics but long-range Coulomb forces, such as electrons trapped in a quantum dot embedded in a thin quantum wire, are Wigner molecules at similar coupling constants. It will be useful to study in the future the intermediate case of dipolar interactions, which are attracting so much interest in the cold-atom literature~\cite{lahaye_dipolargases,steffiWM,xu_arXiv_2011}. Since one-dimensional few-fermion systems with tunable short-range interactions have been recently created~\cite{serwane_science_2011}, our predictions can be tested experimentally by using local probes that access both spatial density distribution and density-density correlation functions.

\begin{acknowledgements} 
Gao X. is supported by NSFC and Zhejiang Provincial NSFC under Grant Nos. 11174253 and R611015. M. R. thanks S. Reimann for useful discussions and acknowledges support from Fondazione Cassa di Risparmio di Modena through the project COLDandFEW and from CINECA-ISCRA through grant no. HP10C1E8PI. 
\end{acknowledgements}

\end{document}